\documentclass[a4paper]{jpconf}
\usepackage{graphicx}
\newcommand       \Angstrom     {\,{\rm \AA}}     
\newcommand       \simlt        {\leq}
\newcommand       \simgt        {\geq}
\newcommand       \gtsim        {\geq}
\newcommand       \ltsim        {\leq}
\newcommand       \magkpc       {\,{\rm mag\, kpc}^{-1}}
\newcommand       \pc           {\,{\rm pc}}

\newcommand       \g            {\,{\rm g}}

\newcommand       \K            {\,{\rm K}}

\newcommand       \ppm          {\,{\rm ppm}}
\newcommand       \um           {\,{\rm \mu m}}
\newcommand       \mum           {\,{\rm \mu m}}
\newcommand       \micron       {\,{\rm \mu m}}
\newcommand       \umrev        {\,{\rm \mu m^{-1}}}
\newcommand       \cm           {\,{\rm cm}}

\newcommand       \lambdamax    {\lambda_{\rm  max}}
\newcommand       \Pmax         {P_{\rm  max}}

\newcommand       \magni        {\,{\rm mag}}
\newcommand       \url          {\,{\rm http://}}
\newcommand       \simali       {{\sim\,}}
\def	\GHz    {\,{\rm GHz}}
\def    \NH     {N_{\rm H}}

\def    \ltsim  {<}
\def    \gtsim  {>}
\def    \simlt  {<}
\def    \simgt  {>} 

\def	  \csun         {\left[{\rm C/H}\right]_{\odot}}

\def	  \osun         {\left[{\rm O/H}\right]_{\odot}}

\begin{document}
\title{Interstellar Grains -- The 75$^{\rm TH}$ Anniversary}

\author{Aigen Li}

\address{Department of Physics and Astronomy,
         University of Missouri, Columbia, MO 65211, USA}
\ead{LiA@missouri.edu, http://bengal.missouri.edu/$\sim$lia/}

\vspace*{-11.2em}
{\noindent \normalsize\tt invited review article for 
the {\bf ``Light, Dust and Chemical Evolution''} conference
(Gerace, Italy, 26--30 September 2004), edited by F. Borghese 
and R. Saija, {\it Journal of Physics: Conference Series}, 2005, 
in press}
\vspace*{7.5em}

\begin{abstract}
The year of 2005 marks the 75th anniversary 
since Trumpler (1930) provided the first definitive 
proof of interstellar grains by demonstrating 
the existence of general absorption and reddening 
of starlight in the galactic plane.
This article reviews our progressive understanding of 
the nature of interstellar dust.
\end{abstract}

\vspace{-12mm}
\section{Introduction: A Brief History for the Studies of 
Interstellar Dust}
In 1930 -- exactly 75 years ago, the existence of solid dust 
particles in interstellar space was by the first time firmly
established, based on the discovery of color excesses 
(Trumpler 1930). But the history of the interstellar dust-related 
studies is a much longer and complex subject, and can be dated 
back to the late 18th century when Herschel (1785) described 
the dark markings and patches in the sky as ``holes in the heavens''. 
Below is a summary of the highlights of this history.
For a more detailed record of the historical development 
of dust astronomy, I refer the interested readers to 
Aiello \& Cecchi-Pestellini (2000), Dorschner (2003), 
Li \& Greenberg (2003), and Verschuur (2003).
\begin{center}
\underline{\bf Early Chronology: From ``Holes in the Heavens''}\\ 
\underline{\bf to the Firm Establishment of Interstellar Dust}
\end{center}
\vspace{-3mm}
\begin{itemize}
\item As early as 1785, Sir William Herschel noticed that the sky
looks patchy with stars unevenly distributed and some regions are 
particularly ``devoid'' of stars. He described these dark regions 
(``star voids'') as ``{\bf holes in the heavens}''. 
\item At the beginning of the 20th century, astronomers started to 
recognize that the ``starless holes'' were real physical structures
in front of the stars, containing {\bf dark obscuring masses of matter} 
able to absorb starlight (Clerke 1903; Barnard 1919), 
largely thanks to the new technology of photography 
which made the photographic survey of the dark markings possible. 
Sir Harold Spencer Jones (1914) also attributed the dark lanes 
seen in photographs of edge-on spiral galaxies to obscuring matter.
Whether the dark lanes in the Milky Way is caused by obscuring material 
was one of the points of contention in the Curtis-Shapley debate 
(Shapley \& Curtis 1921). 
\item Wilhelm Struve (1847) noticed that the apparent number of 
stars per unit volume of space declines in all directions receding 
from the Sun. He attributed this effect to 
{\bf interstellar absorption}.\footnote{%
  To be precise, this should be called 
  ``extinction'' which is a combined effect of absorption 
  and scattering: a grain in the line of sight between a distant 
  star and the observer reduces the starlight by a combination of 
  scattering and absorption.
  }
From his analysis of star counts he deduced an visual extinction
of $\simali 1\magkpc$. Many years later, Jacobus Kapteyn (1904)
estimated the interstellar absorption to be $\simali 1.6\magkpc$,
in order for the observed distribution of stars in space to be 
consistent with his assumption of a constant stellar density. 
This value was amazingly close to the current estimates of 
$\simali 1.8\magkpc$. Max Wolf (1904) demonstrated the existence
of discrete clouds of interstellar matter by comparing the star
counts for regions containing obscuring matter with those for
neighbouring undimmed regions.    
\item In 1912, Vesto Slipher discovered {\bf reflection nebulae} 
from an analysis of the spectrum of the nebulosity
in the Pleiades cluster which he found was identical to that of
the illuminating stars. It was later recognized that the nebulosity
was created by the scattering of light from an intrinsically 
luminous star by the dust particles in 
the surrounding interstellar medium (ISM). 
\item Henry N. Russell (1922) argued that {\bf dark clouds 
accounted for the obscuration and this obscuring matter 
had to be in the form of millimeter-sized fine dust}. 
Anton Pannekoek (1920) recognized that the obscuration 
can not be caused by the Rayleigh scattering of gas,
otherwise one would require unrealistically high masses
for the dark nebulae. He also noticed that, as suggested
by Willem de Sitter, the cloud mass problem can be vanished
if the extinction is due to dust grains with a size comparable 
to the wavelength of visible light.  
\item In 1922, Mary L. Heger observed two broad absorption 
features at 5780$\Angstrom$ and 5797$\Angstrom$, conspicuously 
broader than atomic interstellar absorption lines. 
The interstellar nature of these absorption features was
established 12 years later by Paul W. Merrill (1934). 
These mysterious lines -- known as the {\bf diffuse interstellar 
bands (DIBs)}, still remain unidentified. 
\item In 1930, a real breakthrough was made by 
{\bf Robert J. Trumpler who provided the first unambiguous 
evidence for interstellar absorption and reddening
which led to the general establishment of the existence 
of interstellar dust}. 
Trumpler (1930) based this on a comparison between
the photometric distances and geometrical distances 
of 100 open clusters.\footnote{%
  The photometric distances were 
  obtained by comparing apparent and absolute magnitudes, 
  with the latter determined from the spectral types of 
  the stars in the clusters. The geometrical distances were 
  determined from the angular diameters of the clusters, 
  assuming that all their diameters were the same.
  } 
If there was no interstellar absorption, the two distances 
should be in agreement. However, Trumpler (1930) found that 
the photometric distances are systematically larger than 
the geometrical distances, indicating that the premise
of a transparent ISM was incorrect.\footnote{%
  As mentioned earlier in this review, 
  general star counts did suggest the existence of
  interstellar extinction which increases with distance.
  However, this evidence is not decisive because
  interpretation of the star-count data rests on
  assumptions (generally unproved at the time) 
  as to the true spatial distribution of the stars. 
  }
Using this direct and compelling method he was able to 
find both absorption and selective absorption or color 
excess with increasing distance.\footnote{%
  Trumpler (1930) derived a color-excess of
  $\simali$0.3$\magkpc$ between the photographic 
  (with an effective wavelength 
  $\lambda_B$\,$\approx$\,$4300\Angstrom$) and visual 
  ($\lambda_V$\,$\approx$\,$5500\Angstrom$) bands,
  and a general (visual) absorption of $\simali$1.0$\magkpc$.
  }
Trumpler (1930) also concluded that {\bf the observed
color excess could only be accounted for by ``fine
cosmic dust''}.
\item In 1932, Jan H. Oort demonstrated that the space 
between the stars must contain a considerable amount 
of matter. He derived an {\bf upper limit 
(``Oort limit'') on the total mass of the matter 
(including both stars and interstellar matter)} in 
the solar neighbourhood from an analysis of the motions 
of K giants perpendicular to the plane of the Galaxy 
(the $z$-direction).
An upper limit of $\simali$$1.0\times 10^{-23}\g\cm^{-3}$
on the total mass density was obtained from measuring 
the gravitational acceleration in the $z$-direction.
The Oort limit has important implications:
(1) {\bf there has to be more material in the galactic
plane than could be seen in stars} since the mass density 
of known stars is only $\simali$$4.0\times 10^{-24}\g\cm^{-3}$;
and (2) {\bf the upper limit of $\simali$$6.0\times 10^{-24}\g\cm^{-3}$
on the mass density of the interstellar matter in the solar 
neighbourhood places severe restrictions on the source
of the obscuration}: what kind of material distributed 
with this density with what mass absorption coefficient could 
give rise to the observed visual extinction of about $1\magkpc$?
apparently, only with small dust grains could so much extinction 
by so little mass (and the $\lambda^{-1}$ wavelength dependence; 
see below) be explained. 
\begin{center}
\underline{\bf Interstellar Absorption and Scattering}
\end{center}
\vspace{-2mm}
\item In 1936, Rudnick by the first time measured 
the wavelength dependence of extinction in the wavelength 
range 4000--6300$\Angstrom$ based on differential 
spectrophotometric observations of reddened and unreddened stars 
of the same spectral type. Rudnick (1936) found that the measured 
{\bf extinction curve was inconsistent with Rayleigh scattering} 
(which has a $\lambda^{-4}$ wavelength dependence).
This so-called ``{\bf pair-match}'' method remains the most
common way of deriving an interstellar extinction curve.
\item By the end of the 1930s, a {\bf $\lambda^{-1}$ extinction law 
in the wavelength range 1--3$\umrev$} had been well established 
(Hall 1937; Greenstein 1938; Stebbins, Huffer, \& Whitford 1939),
thanks to the advent of the photoelectric photometry,
excluding free electrons, atoms, molecules, 
and solid grains much larger or much smaller than 
the wavelength of visible light,
leaving solids with a size comparable to the wavelength
as the sole possibility.
\item In 1936, Struve \& Elvey demonstrated the scattering of 
general starlight by interstellar clouds based on a series of 
observations of the dark cloud Barnard~15, the core of which
is appreciably darker than the rim, although the latter is about 
opaque as the former. They attributed the increased brightness
of the outer region to {\bf interstellar scattering}.
\item In 1941, Henyey \& Greenstein confirmed the existence of
{\bf diffuse interstellar radiation} (which was originally detected 
by van Rhijn [1921]) in the photographic wavelength region. 
They interpreted the observed intensity of diffuse light as 
scattered stellar radiation by {\bf interstellar grains which 
are strongly forward scattering and have a high albedo
(higher than $\simali$0.3)}.
\item In 1943, with the advent of the six-colour photometry
(at 3530$\Angstrom$\,$<$\,$\lambda$\,$<$\,10300$\Angstrom$)
Stebbins \& Whitford found that the extinction
curve exhibits curvature at the near infrared (IR;
$\lambda \approx 1.03\,\mu$m) and ultraviolet 
(UV; $\lambda \approx 0.35\,\mu$m) regions,
{\bf deviating from the simple $\lambda^{-1}$ law}.
\item In 1953, Morgan, Harris, \& Johnson estimated
the ratio of total visual extinction to color excess
to be $A_V/E(B-V)\approx 3.0\pm 0.2$. This was supported 
by a more detailed study carried out by Whitford (1958),  
who argued that there appeared to be 
a ``very close approach to uniformity of the reddening
law'' in most directions.
{\bf A uniform extinction curve with a constant $A_V/E(B-V)$}
was welcomed by the astronomical community -- at that
early stage, interstellar dust was mainly regarded as
an annoying mere extinguisher of starlight which prevented
an accurate measurement of distances to stars.
The proposal of ``a uniform extinction curve with 
a constant $A_V/E(B-V)$'' made it easier to correct 
photometric distances for the effects of absorption
(also because the determination of the color excess $E(B-V)$ 
for early-type stars was relatively straightforward).
\item In 1955, based on the UBV photometry of early O stars
in a region in Cygnus, Johnson \& Morgan noted that there may
exist {\bf regional variations} in the interstellar extinction curve. 
The {\bf nonuniformity nature of the interstellar extinction curve}
was later confirmed in Cygnus, Orion, Perseus, Cepheus 
and NGC\,2244 by Johnson \& Borgman (1963), Nandy (1964),
and Johnson (1965). Those authors also found a wide variety
of $A_V/E(B-V)$ values (ranging from $\simali$3.0 to $\simali$7.4)
in different regions. Wampler (1961) found a systematic variation 
with galactic longitude of $E(U-B)/E(B-V)$, the ratio of slopes 
in the blue to those in the visible region.
\item In the 1960s and early 1970s,
the extension of the extinction curve toward the middle 
and far UV ($\lambda^{-1}$$\gtsim$$3\mum^{-1}$) 
was made possible by rocket and satellite observations,
including the rocket-based photoelectric 
photometry at $\lambda=2600\Angstrom$ and 2200$\Angstrom$
(Boggess \& Borgman 1964); the {\it Aerobee} rocket spectrophotometry
at 1200$\Angstrom$\,$\ltsim$\,$\lambda$\,$\ltsim$3000$\Angstrom$ 
(Stecher 1965);
the {\it Orbiting Astronomical Satellite} 
(OAO-2) spectrophotometry at 
1100$\Angstrom$\,$\ltsim$\,$\lambda$\,$\ltsim$\,$3600\Angstrom$ 
(Bless \& Savage 1972);
and the {\it Copernicus} Satellite spectrophotometry at
1000$\Angstrom$\,$\ltsim$\,$\lambda$\,$\ltsim$\,$1200\Angstrom$ 
(York et al.\ 1973).
{\bf By 1973, the interstellar extinction curve had been 
determined over the whole wavelength range from 
0.2$\mum^{-1}$ to 10$\mum^{-1}$}.
\item In 1965, the {\bf 2175$\Angstrom$ extinction bump was detected}
by Stecher (1965). Shortly after its detection, it was attributed
to graphite (Stecher \& Donn 1965).\footnote{%
  The exact nature of the carrier of this bump remains 
  unknown. It is generally believed to be caused by aromatic 
  carbonaceous (graphitic) materials, very likely a cosmic mixture
  of polycyclic aromatic hydrocarbon (PAH) molecules
  (Joblin, L\'eger \& Martin 1992; Li \& Draine 2001b).
  }
It was later found that the strength and width of this bump
vary with environment while its peak position is quite invariant.
\item Cardelli, Clayton, \& Mathis (1989) found that
the optical/UV extinction curve in the wavelength 
range of 0.125$\le$$\lambda$$\le$$3.5\mum$
which shows considerable regional variations 
can be approximated by an analytical formula involving 
only one free parameter: the total-to-selective extinction
ratio $R_V$$\equiv$$A_V/E(B-V)$, 
whereas the near-IR extinction curve 
(0.9$\mum$$\le$$\lambda$$\le$$3.5\um$) can 
be fitted reasonably well by a power law 
$A(\lambda)$$\sim$$\lambda^{-1.7}$, 
showing little environmental variations.\footnote{%
  Very recently, on the basis of the {\it FUSE} 
  observations of 9 Galactic sightlines at
  $1050\Angstrom < \lambda < 1200\Angstrom$,
  Sofia et al.\ (2005) found that the CCM prediction
  for short-wavelengths ($\lambda^{-1}>8\mum^{-1}$) is
  not valid for all sightlines.
  }
\begin{center}
\underline{\bf From Metallic Grains to Dirty Ices:}\\
\underline{\bf Meteoritic Origin or Interstellar Condensation?}
\end{center}
\vspace{-2mm}
\item In the 1930s, small {\bf metallic particles} were proposed 
to be responsible for the interstellar extinction,
partly because meteoritic particles (predominantly metallic) 
and interstellar grains were then thought to have the same origin.
Reasonably good fits to the $\lambda^{-1}$ extinction law 
were obtained in terms of small metallic grains with a dominant 
size of $\simali$0.05$\mum$ (Schal\'{e}n 1936) or a power-law size 
distribution $dn(a)/da$\,$\simali$$a^{-3.6}$ in the size range
$80\Angstrom$\,$\ltsim$\,$a$\,$\ltsim$\,$1\cm$ (Greenstein 1938). 
\item In 1935, based on the correlation between gas concentration 
and extinction, Bertil Lindblad suggested that {\bf interstellar
grains were formed by condensation from the interstellar gas
through random accretion of gas atoms}, as speculated by 
Sir Arthur Eddington (1926) that it was so cold in space 
that virtually all gaseous atoms and ions which hit a solid 
particle would freeze down upon it.\footnote{%
  Van de Hulst (1949) pointed out that this is not
  the case for H, He and Ne since they will evaporate rapidly
  at grain temperatures exceeding $\simali$5$\K$.
  }
However, it was found later that in typical interstellar 
conditions, the Lindblad condensation theory would result in 
a complete disappearance of all condensable gases and the grains 
would grow to sizes ($\simali$10$\mum$) well beyond those which 
could account for the interstellar extinction.
\item In 1946, by introducing a grain destruction process 
caused by grain-grain collisions as a consequence of interstellar 
cloud encounters, Jan H. Oort and Hendrik C. van de Hulst further 
developed the interstellar condensation theory and led to 
the {\bf ``dirty ice'' model consisting of saturated molecules} 
such as H$_2$O, CH$_4$, and NH$_3$ with an equilibrium size 
distribution which could be roughly approximated by 
a functional form    
$dn(a)/da\,$$\simali$$\exp\left[-5 \left(a/0.5\,\mu {\rm m}\right)^3\right]$
and an average size of $\simali$0.15$\mum$. 
What might be the condensation nulcei was unclear at that time. 
\item In 1946, van de Hulst by the first time made 
{\bf realistic estimations of 10--20$\K$ for grain 
temperatures}. Before that, it was long thought
that they had a black-body temperature of $\simali$3.2$\K$
(Eddington 1926). Van de Hulst (1946) noted that interstellar
grains are much warmer than a 3.2$\K$ black-body because
they do not radiate effectively at long wavelengths.
\begin{center}
\underline{\bf Interstellar Polarization}
\end{center}
\vspace{-2mm}
\item In 1949, Hall and Hiltner independently discovered 
the general {\bf interstellar linear polarization} by incident
-- their original objective was to look for intrinsic 
stellar polarization from eclipsing binaries.  
The interstellar origin of this polarization was indicated
by the correlation of the degree of polarization with reddening 
and the fact that the direction of polarization is generally
parallel to the galactic plane. 
The interstellar polarization was attributed to
the {\bf differential extinction of starlight by nonspherical
grains aligned to a small degree with respect to the galactic plane}. 
\item In 1951, Davis \& Greenstein suggested that interstellar
grains could be aligned with respect to the interstellar
magnetic field by the paramagnetic relaxation mechanism.
\item The variation of interstellar polarization with wavelength 
was first revealed by Behr (1959) and Gehrels (1960).
It was later shown that the {\bf wavelength dependence of polarization}
is well approximated by an empirical formula, often known as 
the {\bf Serkowski law} (Serkowski 1973; Coyne, Gehrels, 
\& Serkowski 1974; Wilking et al.\ 1980).\footnote{%
  The ``Serkowski law'' 
  $P(\lambda)/\Pmax$\,=\,$\exp\,[-K\ln^2(\lambda/\lambdamax)]$
  is determined by only one parameter:
  $\lambdamax$ -- the wavelength where the maximum polarization 
  $\Pmax$ occurs; the width parameter $K$ is related
  to $\lambdamax$ through $K$$\approx$$1.66 \lambdamax$\,+\,0.01.
  The peak wavelength $\lambdamax$ is indicative of grain size 
  and correlated with $R_V$: $R_V \approx (5.6\pm 0.3)\lambdamax$ 
  ($\lambdamax$ is in micron; see Whittet 2003).
  }
But the near-IR (1.64$\mum$$<$$\lambda$$<$$5\mum$) 
polarization is better approximated by 
a power law $P(\lambda)$\,$\propto$\,$\lambda^{-\beta}$,
with $\beta$$\simeq$$1.8\pm0.2$, 
independent of $\lambda_{\rm max}$
(Martin \& Whittet 1990, Martin et al.\ 1992).
\item In 1972, the interstellar {\bf circular polarization} which
arises from the interstellar birefringence (Martin 1972) as 
originally predicted by van de Hulst (1957), was first detected 
along the lines of sight to the Crab Nebula by Martin, Illing, 
\& Angel (1972) and to six early-type stars by 
Kemp \& Wolstencroft (1972).  
\begin{center}
\underline{\bf From Dirty Ices to Graphite: Interstellar Condensation 
or Stellar Origin?}
\end{center}
\vspace{-2mm}
\item In early 1950s -- soon after the discovery of interstellar
polarization, the validity of the ice model seemed doubtful since
{\bf ice grains are an inefficient polarizer}, and therefore it 
would be difficult for them to explain the observed rather high 
degree of polarization relative to extinction (van de Hulst 1950; 
Spitzer \& Tukey 1951; Cayrel \& Schatzman 1954).
\item In 1954, Cayrel \& Schatzman suggested that {\bf graphite grains},
comprising a small component of the total mass of interstellar
dust, could account for the observed polarization-to-extinction 
ratio because of their strong optical anisotropy.
\item In 1962, Hoyle \& Wickramasinghe proposed that {\bf graphite grains 
of sizes a few times 0.01$\mum$ could condense in the atmospheres of 
cool N-type carbon stars}, and these grains will subsequently be driven 
out of the stellar atmospheres and {\bf injected into interstellar space} 
by the stellar radiation pressure. Hoyle \& Wickramasinghe (1962)
argued that $\simali$10$^4$ N-type stars in the Galaxy may be 
sufficient to produce the required grain density to
account for the observed interstellar extinction.
They also showed that the extinction predicted from small graphite 
grains is in remarkable agreement with the observed reddening
law (which was then limited to $\lambda^{-1}$\,$<$\,3$\mum^{-1}$).    
It is interesting to note that the condensation of graphite
grains in cool carbon stars was suggested many years earlier
by O'Keefe in 1939, while as early as 1933 Wildt had already found 
that solid grains of carbon, Al$_2$O$_3$, CaO, carbides (SiC, TiC, ZrC),
and nitrides (TiN, ZrN) might form in N-type stars. 
\item In 1966, in view of the fact that the albedo of pure graphite 
grains appear to be too low to be consistent with the observations,
Wickramasinghe, Dharmawardhana, \& Wyld proposed that interstellar 
dust consists of graphite cores and ice mantles.
Wickramasinghe (1965) argued that graphite grains ejected from 
stars tend to grow ice mantles in interstellar clouds. 
Wickramasinghe et al.\ (1966) showed that graphite grains 
of radii $\simali$0.05--0.07$\mum$ coated by an ice mantle 
up to twice their radii could satisfy the observed interstellar
extinction and albedo.
\item In 1968, Wickramasinghe \& Nandy argued that
solid molecular hydrogen mantles may be accreted by 
interstellar grains in dense interstellar clouds. 
Wickramasinghe \& Krishna Swamy (1969) showed that
graphite core-solid H$_2$ mantle grains with core
radii $\simali$0.04--0.06$\mum$ and mantle radii
$\simali$0.15--0.25$\mum$ are consistent with 
the observed interstellar extinction in the wavelength
range of $0.11\mum$\,$\simlt$\,$\lambda$\,$\simlt$\,2$\mum$
and the albedo and phase function derived from the
diffuse Galactic light.
\begin{center}
\underline{\bf Interstellar Silicate Dust of Stellar Origin}
\end{center}
\vspace{-2mm}
\item In 1963, Kamijo suggested that nanometer-sized {\bf SiO$_2$ grains 
could condense in the atmospheres of cool M-type stars}. After blown 
out of the stellar atmospheres and {\bf injected into interstellar space}, 
they could serve as condensation nuclei for the formation of 
``dirty ices''. 
\item In 1968, Wickramasinghe \& Krishna Swamy considered 
quartz grains covered with dirty ice mantles and found 
that their match to the observed extinction curve was 
unsatisfactory.
\item In 1969, Gilman found that {\bf grains around oxygen-rich cool 
giants are mainly silicates} such as Al$_2$SiO$_5$ and Mg$_2$SiO$_4$. 
Silicates were first detected in emission in M stars 
(Woolf \& Ney 1969; Knacke et al.\ 1969a).
\begin{center}
\underline{\bf Interstellar Iron, SiC, and Diamond Grains 
of Stellar Origin?}
\end{center}
\vspace{-2mm}
\item In 1965, Cernuschi, Marsicano, \& Kimel argued that 
{\bf iron grains} could condense out of the expanding 
{\bf supernova explosion ejecta}. Schal\'{e}n (1965) explicitly 
modeled the interstellar extinction curve in the wavelength range 
of $0.5\mum^{-1} < \lambda^{-1} < 4.5\mum^{-1}$
using iron grains of radii $\simali$0.01$\mum$.
Hoyle \& Wickramasinghe (1970) also argued that a significant 
fraction of the mass of the heavy elements produced in supernova 
explosion could condense into solid particles during the expansion 
phase following explosion. They further suggested that supernovae 
may constitute a major source of silicate, iron, and graphite 
grains in the ISM.\footnote{%
  Many years later, the idea of metallic iron grains as an 
  interstellar dust component was reconsidered by Chlewicki \& 
  Laureijs (1988) who attributed the 60$\mum$ emission 
  measured by IRAS for the Galactic diffuse ISM to small 
  iron particles with a typical size of $a$$\approx$70$\Angstrom$
  (which would obtain an equilibrium temperature of $\simali$53$\K$
  in the diffuse ISM). But their model required almost all cosmic 
  iron to be contained in metallic grains: $\simali$34.5\,ppm 
  (parts per million) relative to H.
  Exceedingly elongated metallic needles with a length ($l$) 
  over radius ($a$) ratio $l/a \approx 10^5$, presumably present
  in the intergalactic medium, have been suggested  
  by Wright (1982), Hoyle \& Wickramasinghe (1988), 
  and Aguirre (2000) as a source of starlight opacity
  to thermalize starlight to generate the microwave background. 
  Very recently, elongated needle-like metallic grains were
  suggested by Dwek (2004) as an explanation for 
  the flat 3--8$\mum$ extinction observed by Lutz et al.\ (1996)
  toward the Galactic Center and by Indebetouw et al.\ (2005) 
  toward the $l$=42$^{\rm o}$ and 284$^{\rm o}$ lines of sight
  in the Galactic plane. But these results heavily rely on
  the optical properties of iron needles (see Li 2003a, 2005b).
  }
\item In 1969, Friedemann showed that {\bf silicon carbide grains} 
could condense in the atmospheres of carbon stars and then leave 
the star and become an interstellar dust component, 
although they comprise only a minor fraction of 
the total interstellar dust mass.\footnote{%
  Whittet, Duley, \& Martin (1990) estimated
  from the 7.7--13.5$\mum$ spectra (with a spectral resolution
  of $\simali$0.23$\mum$) of 10 sightlines
  toward the Galactic Center the abundance of Si in SiC dust 
  to be no more than $\simali$5\% of that in silicates.
  Since about half of the dust in the ISM is injected by 
  carbon stars in which an appreciable fraction of the stardust
  is SiC, it is unclear how SiC is converted to gas-phase 
  and recondense to form silicates in the ISM. 
  }
\item In 1969, Saslaw \& Gaustad suggested that carbon may condense
in cool stellar atmospheres in the form of {\bf diamond grains} and 
are subsequently injected into interstellar space.\footnote{%
  Nanodiamonds were identified in
  the dust disks or envelopes surrounding
  two Herbig Ae/Be stars HD 97048 and Elias 1
  and one post-asymptotic giant branch (AGB) star HR 4049,
  based on the 3.43$\mum$ and 3.53$\mum$ 
  C--H stretching emission features 
  expected for surface-hydrogenated nanodiamonds
  (Guillois, Ledoux, \& Reynaud 1999; 
  van Kerckhoven, Tielens, \& Waelkens 2002). 
  }
Presolar nanodiamonds were first detected in primitive carbonaceous 
meteorites based on their isotopic anomalies (Lewis et al.\ 1987;
see \S5.4 of Li \& Draine 2004a and Jones \& d'Hendecourt 2004
for more information regarding interstellar nanodiamonds).
\begin{center}
\underline{\bf Grain Mixtures with Multi-modal Size Distributions}
\end{center}
\vspace{-2mm}
\item The extension of the wavelength base for the
interstellar extinction observations into the
far-UV and IR provide a strong stimulus for
the development of dust models. 
The fact that the extinction continues to increase 
in the far UV (e.g., see York et al.\ 1973)
implies that {\bf no single grain type with either a single size 
or a continuous size distribution could account for the observed 
optical to far-UV interstellar extinction} (Greenberg 1973). 
This led to the abandonment of any one-component grain models 
and stimulated the emergence of various kinds of models 
consisting of multiple dust constituents,
including silicate, SiC, iron, iron oxide, 
graphite, dirty ice, solid H$_2$, etc.\footnote{%
  The reason why so many different materials with such 
  a wide range of optical properties could be used to
  explain the observed interstellar extinction 
  was that the number of free parameters defining the size
  distribution was sufficiently large.
  }
{\bf By early 1970s}, the two highly refractory components
-- {\bf silicates and graphite have been considered in most
dust models}, supported by the detection of the conspicuous
bump in the interstellar extinction curve at 2175$\Angstrom$
and the prominent emission feature at 10$\mum$
of oxygen-rich stars and by the belief that graphite and
silicate grains can be produced in stellar atmospheres and
expelled into the ISM.
\begin{itemize}
\item In 1969, Hoyle \& Wickramasinghe
modeled the interstellar extinction in terms of a mixture
of {\bf silicate} grains of radii $\simali$0.07$\mum$ and 
{\bf graphite} grains of radii $\simali$0.065$\mum$.
\item Wickramasinghe (1970a) found that 
the interstellar extinction curve in the wavelength
range 0.3$\simlt$\,$\lambda^{-1}$\,$\simlt$\,9$\mum^{-1}$ 
could be reproduced by a mixture of {\bf graphite} grains 
with a size distribution of
$dn(a)/da\,$$\simali$$\exp\left[-0.5 \left\{\left(a-0.06\right)/0.02\right\}^2\right]$ for 0.03\,$\mum$$\simlt$\,$a$\,$\simlt$\,0.13$\mum$ 
and {\bf silicate} grains of radii $\simali$0.07$\mum$.
He also found that {\bf silicate} grains of radii $\simali$0.03$\mum$
with an {\bf ice mantle} of radii $\simali$0.14$\mum$ together with
the same graphite population could fit the observed extinction
curve equally well.\footnote{%
  The reason why Wickramasinghe (1970a) considered ice-coated silicate
  grains was that he thought that graphite grains of a typical size
  $\simali$0.06$\mum$ would attain an equilibrium temperature
  of $\simali$40$\K$ in the ISM and would be too warm to possess
  an ice mantle, while silicate grains would tend to take up
  lower temperatures because of their lower optical and UV 
  absorptivity and therefore the condensation of ice mantles
  could occur on their surfaces.
  } 
By modeling the albedoes and phase functions derived from
the diffuse Galactic light, Wickramasinghe (1970b)
concluded that the graphite-silicate mixture was preferred
over the graphite-(ice-coated) silicate mixture. 
\item Wickramasinghe \& Nandy (1970)
found that a mixture of {\bf silicate, graphite, and iron
grains} also achieved a rough fair fit to the interstellar
extinction curve at $\lambda^{-1} \simlt 8\mum^{-1}$.
\item Huffman \& Stapp (1971) found that {\bf enstatite} grains 
plus 12\% small ($\simali$100$\Angstrom$) {\bf iron oxide} grains 
also provided a fairly good fit to the extinction curve up to
$\lambda^{-1} \simlt 8\mum^{-1}$.
\item Gilra (1971) performed extinction calculations
for a mixture of {\bf graphite, silicate, and SiC} and
provided close fits to the observed extinction
curves. But his model heavily relied on SiC:
the required mass of the SiC component was 
$\simali$4 times of that of graphite.
\item Greenberg \& Stoeckly (1970) found that
{\bf ice-coated cylindrical silicate grains} together with 
a population of {\bf small bare silicate grains} could reproduce
the extinction curve from the IR to the UV and 
the wavelength dependence of polarization.
\item In 1974, Greenberg \& Hong suggested that interstellar 
grains consist of {\bf submicron-sized silicate cores 
surrounded by mantles of heterogeneous molecular and 
free-radical mixture of O, C, N and H} (``modified dirty ices''),
and a minor component of {\bf very small bare grains}
of sizes $<$100$\Angstrom$ whose precise composition
was uncertain.
\end{itemize}
\item In a study of the scattering properties of interstellar dust 
(albedo and phase function) determined from the 
OAO-2 observations at 
1500$\Angstrom$\,$\ltsim$\,$\lambda$\,$\ltsim$\,$4250\Angstrom$ 
of the diffuse Galactic light (Witt \& Lillie 1973), 
Witt (1973) first explicitly suggested 
a bi-modal size distribution for interstellar grains: 
large grains with radii $\gtsim$$2500\Angstrom$ would provide 
extinction in the visible region including scattering 
which is strongly forward directed, 
and small particles with radii $\ltsim$$250\Angstrom$ would 
dominate the UV region and contribute nearly isotopic scattering.
\begin{center}
\underline{\bf The Infrared Era: Ices, Silicates, PAHs 
and Aliphatic Hydrocarbons}
\end{center}
\vspace{-2mm}
\item In the 1960s, {\bf the first attempt to search for 
the 3.1$\um$ feature of H$_2$O ice in the diffuse ISM was 
unsuccessful} (Danielson, Woolf, \& Gaustad 1965; 
Knacke, Cudaback, \& Gaustad 1969b), although it had long 
been considered to be a possible constituent of interstellar 
grains. This was the strongest objection against 
the dirty-ice model of Oort \& van de Hulst (1946).
\item In 1973, the 3.1$\mum$ H$_2$O ice feature was finally 
detected (Gillett \& Forrest 1973). But it was recognized that 
{\bf water ice is present only in dense regions}
(usually with $A_V$$>$3$\magni$). 
\item By early 1970s, {\bf silicates had been detected in the ISM},
first in emission in the Trapezium region of the Orion Nebula 
(Stein \& Gillett 1969), then in absorption toward the 
Galactic Center (Hackwell, Gehrz, \& Woolf 1970), and toward the
Becklin-Neugebauer object and Kleinmann-Low Nebula (Gillett 
\& Forrest 1973). 
\item In 1973, Gillett, Forrest, \& Merrill (1973) detected 
prominent emission features at 8.6, 11.3, and 12.7$\mum$
in the planetary nebulae NGC~7027 and BD+30$^{\rm o}$3639.
These features together with the 3.3, 6.2, and 7.7$\mum$
features were collectively known as 
the{\bf  ``{\it unidentified infrared}'' (UIR) bands},
which are now often attributed to polycylic aromatic 
hydrocarbon (PAH) molecules (Duley \& Williams 1981;
L\'{e}ger \& Puget 1984; Allamandola, Tielens, \& Barker 1985;
Allamandola, Hudgins, \& Sandford 1999).\footnote{%
  Since the ``UIR'' emission bands were initially found
  to be associated with UV-rich objects, it had been
  thought that they were pumped primarily by UV photons.
  Li \& Draine (2002b) demonstrated that the excitation 
  of PAHs does not require UV photons -- since the PAH 
  electronic absorption edge shifts to longer wavelengths 
  upon ionization and/or as the PAH size increases
  (see Mattioda, Allamandola, \& Hudgins 2005 for
   their recent measurements of the near-IR absorption 
   spectra of PAH ions),
  therefore long wavelength (red and far-red) photons are 
  also able to heat PAHs to high temperatures so that they 
  emit efficiently at the ``UIR'' bands
  (also see Smith, Clayton, \& Valencic 2004).
  Li \& Draine (2002b) have modeled the excitation 
  of PAH molecules in UV-poor regions. 
  It was shown that the astronomical PAH model provides
  a satisfactory fit to the UIR spectrum of vdB\,133, 
  a reflection nebulae with the lowest ratio of UV to total
  radiation among reflection nebulae with detected UIR band 
  emission (Uchida, Sellgren, \& Werner 1998).  
  }
\item Willner et al.\ (1979) detected a strong absorption band
at 3.4$\um$ in the Galactic Center toward Sgr A\,W.
Wickramasinghe \& Allen (1980) detected this feature 
in the Galactic Center source IRS\,7. Although it is generally 
accepted that this feature is due to the C--H stretching mode in 
saturated {\bf aliphatic hydrocarbons}, the exact nature of this 
hydrocarbon material remains uncertain (see Pendleton
\& Allamandola 2002, Pendleton 2004 for recent reviews).
This feature has also been detected in a carbon-rich 
protoplanetary nebula CRL\,618 (Lequeux \& Jourdain de Muizon 1990; 
Chiar et al.\ 1998) with close resemblance to the interstellar 
feature. 
\begin{center}
\underline{\bf Interstellar Depletion: Where Have All Those Atoms Gone?}
\end{center}
\vspace{-2mm}
\item In 1973, Morton et al.\ found that {\bf the gas-phase 
abundances of some heavy elements} (relative to hydrogen)
measured by the {\it Copernicus} UV satellite for 
interstellar clouds are {\bf significantly lower than in the Sun}. 
\item In 1974, Field noted that {\bf the depletions of certain
elements observed by Morton et al.\ (1973) correlate with 
the temperatures for dust condensation} in stellar
atmospheres or nebulae. He suggested that these elements
have condensed into dust grains near stars and other
elements have accreted onto such grains in interstellar 
space after they enter the ISM, forming a mantle composed
of H, C, N and O compounds.
\item In 1974, Greenberg found that the observed depletion 
of C, N, and O is significantly greater than could be accommodated 
by the dust under any reasonable models, using the gas-phase 
abundances measured by {\it Copernicus} for the $\zeta$ Ophiuchi 
sightline  (Morton et al.\ 1973) and the solar abundances 
as the reference abundances.
\item Twenty years later, Sofia, Cardelli, \& Savage (1994) 
found that the interstellar depletions are lowered 
for C, N, and O if B stars are used as the reference standard.
They argued that the solar system may have 
enhanced abundances of many elements, 
and therefore the solar abundances
are not representative of the interstellar abundances.
\item Snow \& Witt (1996) analyzed the surface abundances of
B stars and field F and G stars and found that not only
C, N, and O but also Si, Mg, and Fe and many other elements
are underabundant in these stars. This led them to suggest
that the {\bf interstellar abundances are appreciably 
subsolar} ($\simali$60\%--70\% of the solar values).\footnote{%
  The most recent estimates of the solar C 
  ($\csun \approx 245\ppm$; 
   Allende Prieto, Lambert, \& Asplund 2002) 
  and O abundances ($\osun \approx 457\ppm$; Asplund et al.\ 2004)
  are also ``subsolar'', just $\simali$50\%--70\% of 
  the commonly-adopted solar values 
  (e.g. those of Anders \& Grevesse 1989)
  and close to the ``subsolar'' interstellar abundances
  originally recommended by Snow \& Witt (1996).  
  {\bf If the interstellar abundances are indeed ``subsolar'',
  there might be a lack of raw material to form 
  the dust to account for the interstellar extinction. 
  Mathis (1996) argued that this problem could be solved 
  if interstellar grains have a fluffy, porous structure}
  since fluffy grains are more effective in 
  absorbing and scattering optical and UV starlight than 
  compact grains (on a per unit mass basis). 
  However, {\bf using the Kramers-Kronig relation, Li (2005a) 
  demonstrated that fluffy dust is not able to overcome 
  the abundance shortage problem.}
  The abundances of refractory 
  elements in {\it stellar photospheres}
  may under-represent the composition of the interstellar 
  material from which stars are formed, 
  resulting either from the possible underestimation of
  the degree of heavy-element settling in stellar
  atmospheres, or from the incomplete incorporation
  of heavy elements in stars during the star formation process.
  }
\begin{center}
\underline{\bf Dust Luminescence: The ``Extended Red Emission''}
\end{center}
\vspace{-2mm}
\item In 1980, Schmidt, Cohen, \& Margon (1980), detected in 
the Red Rectangle a far-red continuum emission in excess of 
what would be expected from simple scattering of starlight 
by interstellar dust. This continuum emission, known as 
the ``{\bf extended red emission}'' (ERE), consists of 
a broad, featureless emission band between $\sim$5400$\Angstrom$ 
and 9500$\Angstrom$, 
peaking at $6100\ltsim \lambda_{\rm p} \ltsim 8200\Angstrom$, 
and with a width 
$600\Angstrom\ltsim {\rm FWHM}\ltsim 1000\Angstrom$.\footnote{%
  Very recently, Vijh, Witt, \& Gordon (2004) reported 
  the discovery of blue luminescence 
  at $\lambda$\,$<$\,5000$\Angstrom$ in the Red Rectangle
  and identified it as fluorescence by small three- to four-ringed 
  PAH molecules. Nayfeh, Habbal, \& Rao (2005) argued that
  this blue luminescence could be due to hydrogen-terminated
  crystalline silicon nanoparticles.
  }
The ERE has been seen in a wide variety of dusty environments: 
the diffuse ISM of our Galaxy, reflection nebulae, planetary nebulae, 
HII regions, and other galaxies 
(see Witt \& Vijh 2004 for a recent review). 
\item The ERE is generally attributed to photoluminescence (PL) 
by some component of interstellar dust, 
powered by UV/visible photons. The photon conversion efficiency of
the diffuse ISM has been determined to be near unity
(Gordon, Witt, \& Friedmann 1998).
$\longrightarrow$
{\bf The ERE carriers are very likely in the nanometer size range} 
because nanoparticles are expected to luminesce efficiently 
through the recombination of the electron-hole pair
created upon absorption of an energetic photon,
since in such small systems the excited electron 
is spatially confined and the radiationless transitions 
that are facilitated by Auger and defect related recombination 
are reduced (see Li 2004a).
\item {\bf The ERE carrier remains unidentified}. 
Various candidate materials have been proposed, 
but most of them appear unable to match the observed 
ERE spectra and satisfy the high-PL efficiency requirement 
(Li \& Draine 2002a; Li 2004a; Witt \& Vijh 2004).
Promising candidates include PAHs (d'Hendecourt et al.\ 1986)
and silicon nanoparticles (Ledoux et al.\ 1998, 
Witt, Gordon, \& Furton 1998, Smith \& Witt 2002), 
but both have their own problems
(see Li \& Draine 2002a).  
\begin{center}
\underline{\bf Stochastically Heated Ultrasmall Grains 
or ``Platt'' Particles}
\end{center}
\vspace{-2mm}
\item In 1956, {\bf John R. Platt first suggested that 
very small grains or large molecules of less than 10$\Angstrom$ 
in radius} grown by random accretion from the interstellar gas 
could be responsible for the observed interstellar extinction 
and polarization. Platt (1956) postulated these ``Platt'' particles 
as quantum-mechanical particles containing many ions and 
free radicals with unfilled electronic energy bands.
\item In 1968, Donn further proposed that {\bf PAH-like 
``Platt particles'' may be responsible 
for the UV interstellar extinction.}
%
\item In 1968, Greenberg first pointed out that very small grains with 
a heat content smaller than or comparable to the energy of a single 
stellar photon, cannot be characterized by a steady-state temperature
but rather are subject to {\bf substantial temporal fluctuations 
in temperature}. 
\item Andriesse (1978) by the first time presented observational
evidence for the existence of {\bf ``Platt'' particles in a dust 
cloud near M17}, as indicated by its near-invariant 8--20$\mum$ 
spectral shape over a distance of $\simali$2$^{\prime}$ through 
the source and by its broad spectral energy distribution 
characterized by a combination of widely different color 
temperatures. He found that the the observed IR spectrum 
of M17 could be explained by a population of large grains 
and a population of ``Platt'' particles 
of $\simali$10$\Angstrom$ in size 
which exhibit temperature fluctuations.
\item Sellgren, Werner, \& Dinerstein (1983) found that
the color temperatures of the 2--5$\mum$ near-IR continuum
($\simali$1000\,K) and the spectral shapes of the 3.3$\mum$ 
emission features of three visual reflection nebulae 
NGC 7023, 2023, and 2068 show very little variation from 
source to source and within a given source with distance 
from the central star. They attributed the near-IR continuum 
emission to ultrasmall grains of radii $\simali$10$\Angstrom$ 
undergoing large excursions in temperature due to 
stochastic heating by single stellar photons. 
\item The presence of a population of ultrasmall grains 
in the diffuse ISM was explicitly indicated by 
the 12$\mum$ and 25$\mum$ ``cirrus'' emission 
detected by the {\it Infrared Astronomical Satellite} (IRAS) 
(Boulanger \& P\'{e}rault 1988), which is far in excess 
(by several orders of magnitude) of what would be expected 
from large grains of 15--25$\K$ in thermal equilibrium with 
the general interstellar radiation field. Subsequent measurements 
by the {\it Diffuse Infrared Background Experiment} (DIRBE) 
instrument on the {\it Cosmic Background Explorer} (COBE) satellite 
confirmed this and detected additional broadband emission at 
3.5$\mum$ and 4.9$\mum$ (Arendt et al.\ 1998). 
More recently, spectrometers aboard 
the {\it Infrared Telescope in Space} (IRTS) 
(Onaka et al.\ 1996; Tanaka et al.\ 1996) 
and the {\it Infrared Space Observatory} (ISO) 
(Mattila et al.\ 1996) have shown that the diffuse ISM 
radiates strongly in emission features at 3.3, 6.2, 7.7, 8.6, 
and 11.3$\mum$.
\begin{center}
\underline{\bf Interstellar Grain Models: Modern Era}
\end{center}
\vspace{-2mm}
\item The modern era of interstellar grain models probably began
in 1977 with the paper by Mathis, Rumpl, \& Nordsieck (1977).
By fitting the interstellar extinction over the wavelength range
of 0.11$\mum$\,$\simlt$\,$\lambda$\,$\simlt$\,1$\mum$, 
Mathis et al.\ derived a power-law size distribution of 
$dn/da \sim a^{-3.5}$ for a mixture of 
{\bf bare silicate and graphite grains}.\footnote{%
  Such a power-law size distribution is a natural 
  product of shattering following grain-grain collisions \
  (e.g. see Hellyer 1970, Biermann \& Harwit 1980, Dorschner 1982, 
  Henning, Dorschner, \& G\"urtler 1989).
   } 
With the substantial improvements made 
by Draine \& Lee (1984), this model became one of the standard
interstellar grain models with well-characterized chemical
composition, size distribution, optical and thermal properties.
Modifications to this model were later made by 
Draine \& Anderson (1985), Weiland et al.\ (1986), 
Sorrell (1990), Siebenmorgen \& Kr\"{u}gel (1992), 
Rowan-Robinson (1992), Kim, Martin, \& Hendry (1994),
Dwek et al.\ (1997), Clayton et al.\ (2003),
and Zubko, Dwek, \& Arendt (2003) 
by including new dust components (e.g., amorphous carbon, 
carbonaceous organic refractory, and PAHs) 
and adjusting dust sizes (e.g., deriving dust size
distributions using the ``Maximum Entropy Method''
or the ``Method of Regularization'' rather than
presuming a certain functional form).\\[2mm]
Recent developments were made by Draine and his coworkers 
(Li \& Draine 2001b, 2002b,c; Weingartner \& Draine 2001a) 
who have extended the silicate-graphite grain model 
to explicitly include a PAH component as 
the small-size end of the carbonaceous grain population.
It has been shown that the IR emission spectrum calculated from
this model closely matches that observed for the Milky Way
(Li \& Draine 2001b), 
the Small Magellanic Cloud (SMC; Li \& Draine 2002c),
and more recently the ringed Sb galaxy NGC\,7331
(Regan et al.\ 2004; Smith et al.\ 2004),
including the ``UIR'' emission bands at 3.3, 6.2, 7.7, 8.6,
and 11.3$\micron$.
\item In contrast to the bare silicate-graphite model,
Greenberg (1978) proposed that interstellar grains could
be {\bf coated by a layer of organic refractory material} 
derived from the photoprocessing of ice mantles acquired 
in molecular clouds and repeatedly cycled into and out of 
diffuse and molecular clouds. 
The organic refractory mantles would provide a shield 
against destruction of the silicate cores.
Since the rate of production of silicate dust in stars 
is about 10 times slower than the rate of destruction 
in the ISM (mostly caused by sputtering and grain-grain 
collisions in interstellar shock waves; 
Draine \& Salpeter 1979a,b, Jones et al.\ 1994),
the silicates would be underabundant
if they were not protected and thus it would be hard to 
explain the observed large depletions of Si, Fe and Mg 
and the strength of the observed 9.7$\mum$ silicate absorption 
feature, unless most of the silicate mass 
was condensed in the ISM as suggested by Draine (1990).
The most recent development of this model was that of 
Li \& Greenberg (1997), who modeled the core-mantle grains 
as finite cylinders (to account for the interstellar polarization).
In addition, a PAH component and a population of 
small graphitic grains are added respectively to 
account for the far-UV extinction rise 
plus the ``UIR'' emission bands  
and the 2175$\Angstrom$ extinction bump.\\[2mm]
Modifications to this model were also made by considering 
different coating materials (e.g., amorphous carbon, 
hydrogenated amorphous carbon [HAC]), 
including new dust type (e.g., iron, small bare silicates),
and varying dust size distributions 
(Chlewicki \& Laureijs 1988; Duley, Jones, \& Williams 1989;
D\'{e}sert, Boulanger, \& Puget 1990; Li \& Greenberg 1998;
Zubko 1999). In particular, Duley et al.\ (1989) speculated 
that the silicate cores are coated with a mantle of
HAC material arising from direct accretion of gas-phase 
elemental carbon on the silicate cores in the diffuse ISM. 
\item Recognizing that grain shattering due to grain-grain 
collisions and subsequent reassembly through agglomeration 
of grain fragments may be important in the ISM, 
Mathis \& Whiffen (1989) proposed that interstellar grains 
may consist of a {\bf loosely coagulated structure built up 
from small individual particles of silicates and carbon 
of various kinds} (amorphous carbon, HAC, and organic refractories).
Further developments of this composite model were made
by Mathis (1996), Ita\`i et al.\ (2001, 2004), 
Saija et al.\ (2001, 2003), and Zubko, Dwek, \& Arendt (2003) 
(see \S2 of Li 2004a for more details).
\end{itemize}

\section{Interstellar Grains, What Do We Know?}
Our knowledge of interstellar dust regarding its size, shape 
and composition is mainly derived from its interaction with 
electromagnetic radiation: attenuation (absorption and scattering)
and polarization of starlight, and emission of IR and 
far-IR radiation. Presolar grains identified in
meteorites and interplanetary dust particles (IDPs)
of cometary origin also contain useful information
regarding the nature of interstellar grains.
The principal observational keys, both direct
and indirect, used to constrain the properties of dust were
summarized in recent reviews of Draine (2003) and Li (2004b).
\begin{itemize}
\item {\bf (1) \underline{Grain Sizes}.}~ From the wavelength-dependent 
interstellar extinction and polarization curves as well as the near,
mid and far IR emission, we know that there must exist a distribution
of grains sizes, ranging from a few angstroms to a few micrometers.
\begin{itemize}
\item The interstellar extinction curve contains important
information regarding the grain sizes since generally 
speaking, a grain absorbs and scatters light most effectively 
at wavelengths comparable to its size $\lambda$$\approx$$2\pi a$.
The extinction curve rises from the near-IR to the near-UV, 
with a broad absorption feature at about 
$\lambda^{-1}$$\approx$4.6$\mum^{-1}$ 
($\lambda$$\approx$2175$\Angstrom$),
followed by a steep rise into the far-UV
$\lambda^{-1}$$\approx$10$\mum^{-1}$.
$\longrightarrow$
{\bf There must exist in the ISM a population 
of large grains with $a$$\simgt$$\lambda/2\pi$$\approx$$0.1\mum$
to account for the extinction at visible wavelengths,
and a population of ultrasmall grains 
with $a$$\simlt$$\lambda/2\pi$$\approx$$0.016\mum$
to account for the far-UV extinction at 
$\lambda$=$0.1\mum$} (see Li 2004a for details).
\item The interstellar polarization curve rises from the IR, 
has a maximum somewhere in the optical and then decreases 
toward the UV. $\longrightarrow$ 
{\bf There must exist a population of aligned, nonspherical grains 
with typical sizes of $a$$\approx$$\lambda/2\pi$$\approx$0.1$\mum$ 
responsible for the peak polarization 
at $\lambda$$\approx$0.55$\mum$.}
\item Interstellar grains absorb starlight in the UV/visible 
and re-radiate in the IR. The IR emission spectrum of the Milky
Way diffuse ISM, estimated using the IRAS 12, 25, 60 and 100$\mum$
broadband photometry,
the DIRBE-COBE 2.2, 3.5, 4.9, 12, 25, 60, 100, 140 and
240$\mum$ broadband photometry,
and the FIRAS-COBE 110$\mum$$<$$\lambda$$<$3000$\mum$
spectrophotometry, is characterized by a modified
black-body of $\lambda^{-1.7}B_\lambda$(T=19.5$\K$)
peaking at $\simali$130$\mum$ in the wavelength range
of 80$\mum$$\ltsim$$\lambda$$\ltsim$1000$\mum$,
and a substantial amount of emission at $\lambda$$\simlt$60$\mum$
which far exceeds what would be expected from dust 
at $T$$\approx$20$\K$. In addition, spectrometers aboard 
the IRTS (Onaka et al.\ 1996; Tanaka et al.\ 1996) 
and ISO (Mattila et al.\ 1996) 
have shown that the diffuse ISM radiates strongly in 
emission features at 3.3, 6.2, 7.7, 8.6, and 11.3$\um$. 
$\longrightarrow$ {\bf There must exist a population 
of {\bf ``cold dust''} in the size range of
$a$$>$250$\Angstrom$, heated by starlight 
to equilibrium temperatures of 15$\K$$\simlt$$T$$\simlt$25$\K$}
and cooled by far-IR emission to produce
the emission at $\lambda$$\gtsim$60$\mum$ 
which accounts for $\simali$65\% of the total emitted power
(see Li \& Draine 2001b);
{\bf there must also exist a population 
of {\bf ``warm dust''} in the size range of
$a$$<$250$\Angstrom$, stochastically heated by single
starlight photons to temperatures $T$$\gg$20$\K$}
and cooled by near- and mid-IR emission
to produce the emission at $\lambda$$\ltsim$60$\mum$ 
which accounts for $\simali$35\% of the total emitted power
(see Li \& Draine 2001b; Li 2004a).
\item The scattering properties of dust grains 
(albedo and phase function) provide a means of 
constraining the optical properties of the grains
and are therefore indicators of their size and composition.
The albedo in the near-IR and optical is quite high ($\simali$0.6), 
with a clear dip to $\simali$0.4 around the 2175$\Angstrom$ hump,
a rise to $\simali$0.8 around $\lambda^{-1}$$\approx$6.6$\mum^{-1}$,
and a drop to $\simali$0.3 by $\lambda^{-1}$$\approx$10$\mum^{-1}$;
the scattering asymmetry factor almost monotonically rises from
$\simali$0.6 to $\simali$0.8 from $\lambda^{-1}$$\approx$1$\mum^{-1}$
to $\lambda^{-1}$$\approx$10$\mum^{-1}$ (see Gordon 2004).
$\longrightarrow$ {\bf An appreciable fraction of
the extinction in the near-IR and optical must arise from
scattering; the 2175$\Angstrom$ hump is an absorption feature
with no scattered component; and ultrasmall grains
are predominantly absorptive.} 
\item The ``anomalous'' Galactic foreground microwave emission 
in the 10--100$\GHz$ region (Draine \& Lazarian 1998a,b), 
the photoelectric heating of the diffuse ISM
(Bakes \& Tielens 1994, Weingartner \& Draine 2001b), 
and (probably) the ERE (Witt \& Vijh 2004) also provide
direct or indirect proof for the existence
of {\bf nanometer-sized grains} in the ISM 
(see \S2 in Li 2004a for details).
\item Both {\bf micrometer-sized presolar grains} 
(such as graphite, SiC, corundum Al$_2$O$_3$, and 
silicon nitride Si$_3$N$_4$) and 
{\bf nanometer-sized presolar grains} (such as nanodiamonds 
and titanium carbide nanocrystals)\footnote{%
  von Helden et al.\ (2000) proposed that
  TiC nanocrystals could be responsible for 
  the prominent 21$\mum$ emission feature detected 
  in over a dozen carbon-rich post-AGB stars
  which remains unidentified since its first detection 
  (Kwok, Volk, \& Hrivnak 1989). 
  Based on the Kramers-Kronig relations (Purcell 1969),
  Li (2003b) found that the TiC proposal is not
  feasible because it requires at least 50 times
  more Ti than available.  
  }
of interstellar origin as indicated by their anomalous 
isotopic composition have been identified in primitive 
meteorites (see Clayton \& Nittler 2004 for a recent review). 
Presolar silicate grains have recently been identified 
in IDPs (Messenger et al.\ 2003).
Submicron-sized GEMS (Glass with Embedded Metals and Sulfides) 
of presolar origin have also been identified in IDPs and 
their 8--13$\mum$ absorption spectrum were similar to 
those observed in interstellar molecular clouds and young 
stellar objects (see Bradley 2003 for a recent review).
\item Very large interstellar grains (with radii $a$$>$1$\mum$) 
entering the solar system have been detected by the interplanetary 
spacecraft {\it Ulysses} and {\it Galileo} (Gr\"{u}n et al.\ 1993, 1994). 
Huge grains of radii of $a$$\simali$10$\mum$ whose interstellar
origin was indicated by their hyperbolic velocities
have been detected by radar methods (Taylor et al.\ 1996).
But Frisch et al.\ (1999) and Weingartner \& Draine (2001a)
argued that the amount of very large grains inferred from these
detections were difficult to reconcile with the interstellar 
extinction and interstellar elemental abundances. 
\end{itemize}
\item {\bf (2) \underline{Grain Shape}.}~ The detection of 
interstellar polarization clearly indicates that {\bf some 
fraction of the interstellar grains must be nonspherical 
and aligned.} The fact that the wavelength dependence of 
the interstellar polarization exhibits a steep decrease 
toward the UV suggests that {\bf the ultrasmall grain 
component responsible for the far-UV extinction rise is 
either spherical or not aligned.}
\begin{itemize}
\item The 9.7 and 18$\mum$ silicate absorption features 
are polarized in some interstellar regions, most of which
are featureless.\footnote{%
  The only exception is AFGL 2591, a molecular 
  cloud surrounding a young stellar object, which displays
  a narrow feature at 11.2$\um$ superimposed on the broad 
  9.7$\mum$ polarization band, generally attributed to 
  annealed silicates (Aitken et al.\ 1988).
  However, its 3.1$\mum$ ice absorption feature is not polarized 
  (Dyck \& Lonsdale 1980, Kobayashi et al.\ 1980).
  }
Polarization has also been detected in the 3.1$\um$ 
H$_2$O, 4.67$\um$ CO and 4.62$\um$ OCN$^{-}$
absorption features (e.g. see Chrysostomou et al.\ 1996).
Hough et al.\ (1996) reported the detection of a weak 3.47$\um$ 
polarization feature in the Becklin-Neugebauer object 
in the OMC-1 Orion dense molecular cloud, attributed to 
carbonaceous materials with diamond-like structure.
$\longrightarrow$ The detection of polarization in
both silicate and ice absorption features is consistent
with the assumption of a core-mantle grain morphology
(e.g. see Lee \& Draine 1985).
\item So far only two lines of sight toward HD\,147933 
and HD\,197770 have a weak 2175$\Angstrom$ polarization 
feature detected (Clayton et al.\ 1992; Anderson et al.\ 1996; 
Wolff et al.\ 1997; Martin, Clayton, \& Wolff 1999). 
Even for these sightlines, the degree of alignment 
and/or polarizing ability of the carrier should be very small
(see \S2.1.2.1 in Li \& Greenberg 2003 for details).
$\longrightarrow$ The 2175$\Angstrom$ bump carrier
is a very inefficient polarizer
(i.e. it is either nearly spherical or poorly aligned).  
\item So far, no polarization has been detected for the DIBs 
(see Somerville 1996 for a review), the 3.4$\mum$ absorption 
feature (Adamson et al.\ 1999),\footnote{%
  So far spectropolarimetric measurement of this
  feature has been performed only for one sightline
  -- the Galactic Center source IRS\,7 (Adamson et al.\ 1999).
  Unfortunately, no such measurements have been 
  carried out for the 9.7$\mum$ silicate absorption feature
  of this sightline. Spectropolarimetric measurements for 
  both these two bands of the same sightline would allow
  a direct test of the silicate core-hydrocarbon mantle 
  interstellar dust model (Li \& Greenberg 1997), 
  since this model predicts that the 3.4$\mum$ feature would 
  be polarized if the 9.7$\mum$ feature (for the same sightline) 
  is polarized (Li \& Greenberg 2002).
  } 
and the ``UIR'' emission bands (Sellgren, Rouan, \& L\'eger 1988). 
$\longrightarrow$ Their carriers do not align or lack optical anisotropy.
\end{itemize}
\item {\bf (3) \underline{Grain Composition}.}~ It is now generally 
accepted that interstellar grains consist of amorphous silicates and 
some form of carbonaceous materials; the former is inferred from 
the 9.7$\mum$ Si--O stretching mode and 18$\mum$ O-Si-O bending 
mode absorption features in interstellar regions as well as 
the fact that the cosmically abundant heavy elements such as 
Si, Fe, Mg are highly depleted; the latter is mainly inferred 
from the 2175$\Angstrom$ extinction hump 
(and the ubiquitous 3.4$\mum$ C--H stretching vibrational band) 
and the fact that silicates alone are not able to provide 
enough extinction (see Footnote-14 of Li 2004b). 
\begin{itemize}
\item The 9.7$\mum$ and 18$\mum$ absorption features 
are ubiquitously seen in a wide range of astrophysical 
environments. These features are almost certainly due to
silicate minerals: they are respectively ascribed to 
the Si-O stretching and O-Si-O bending modes in some form 
of silicate material (e.g. olivine Mg$_{2x}$Fe$_{2-2x}$SiO$_4$).
In the ISM, these features are broad and relatively featureless. 
$\longrightarrow$ {\bf Interstellar silicates are largely amorphous 
rather than crystalline.}\footnote{%
  Li \& Draine (2001a) estimated that the {\small amount} of 
  $a$$<$1$\um$ crystalline silicate grains in the diffuse ISM 
  is $<$5\% of the solar Si abundance.
  Kemper, Vriend \& Tielens (2004) placed a much tighter
  upper limit of $\simali$0.2\% on the crystalline 
  fraction of the interstellar silicates along the sightline 
  toward the Galactic Center.
  }
\item The strength of the 9.7$\mum$ feature is approximately
$\Delta \tau_{9.7\mum}/A_V$$\approx 1/18.5$ in the local
diffuse ISM. $\longrightarrow$ {\bf Almost all Si atoms have 
been locked up in silicate dust, if assuming solar abundance 
for the ISM} (see Footnote-9 of Li 2004b).\footnote{%
  The silicate absorption feature (relative to the visual 
  extinction) along the path to the Galactic Center is about 
  twice that of the local ISM: 
  $\Delta \tau_{9.7\mum}/A_V$$\approx 1/9$ (Roche \& Aitken 1985).
  It was originally thought that there were very few carbon
  stars in the central regions of the Galaxy so that one
  would expect a much larger fraction of the dust to be
  silicates than is the case further out in the Galactic 
  disk (Roche \& Aitken 1985).
  However, this explanation was challenged by the fact
  that the 3.4$\mum$ aliphatic hydrocarbon dust absorption
  feature for the Galactic Center sources 
  (relative to the visual extinction:
  $\Delta\tau_{3.4\mum}/A_V$$\approx$1/150) 
  is also about twice that of the local ISM
  ($\Delta\tau_{3.4\mum}/A_V$$\approx$1/250; 
   Pendleton et al.\ 1994; Sandford, Allamandola, \& Pendleton 1995).     
  }
\item The 3.4$\mum$ absorption feature is also ubiquitously 
seen in the diffuse ISM (but never in dense regions) of the Milky Way 
and other galaxies (e.g. Seyfert galaxies and ultraluminous 
infrared galaxies, see Pendleton 2004 for a recent review).
This feature is generally attributed to the C-H stretching 
mode in {\bf aliphatic hydrocarbon dust}, although its exact 
nature remains uncertain.\footnote{%
  Over 20 different candidates have been proposed 
  (see Pendleton \& Allamandola 2002 for a summary). 
  So far, the experimental spectra of hydrogenated
  amorphous carbon (HAC; Schnaiter, Henning \& Mutschke 1999,
  Mennella et al.\ 1999) and the organic refractory residue,
  synthesized from UV photoprocessing of interstellar ice mixtures
  (Greenberg et al.\ 1995), provide the best fit to both the overall 
  feature and the positions and relative strengths of the 3.42$\um$, 
  3.48$\um$, and 3.51$\um$ subfeatures corresponding to symmetric 
  and asymmetric stretches of C--H bonds in CH$_2$ and CH$_3$ groups. 
  Pendleton \& Allamandola (2002) attributed this feature to
  hydrocarbons with a mixed aromatic and aliphatic character. 
  } 
\item In principle, we could estimate the volume ratio of 
the silicate component to the aliphatic hydrocarbon component
(1) if we know the band strength of the carrier of the 3.4$\mum$ 
absorption feature (see Li 2004b), or (2) if we know the total 
abundances of interstellar elements (see Li 2005a). 
However, neither is precisely known.  
\end{itemize}
%
%
\item {\bf (4) \underline{Distribution of Dust 
and its Association with Gas}.}~ 
Interstellar grains are unevenly distributed but 
primarily confined to the galactic plane with an effective 
thickness of $\simali$200$\pc$. On average, the ``rate of 
extinction'' (the amount of visual extinction per unit 
distance) $\langle A_V/L\rangle$ is about $\simali$1.8$\magkpc$ 
for the sightlines close to the galactic plane and for 
distances up to a few kiloparsecs from the Sun (Whittet 2003). 
Assuming a mean grain size of $\simali$0.1$\mum$
and a typical mass density of $\simali$2.5$\g\cm^{-3}$
for the interstellar grain material, we can estimate 
the mean dust number density and mass density in the solar 
neighbourhood ISM respectively to be 
$n_{\rm dust} \approx 1.1\times 10^{-12}\cm^{-3}$
and $\rho_{\rm dust} \approx 1.2\times 10^{-26}\g\cm^{-3}$
from the ``rate of extinction''.\footnote{%
  Let interstellar grains be approximated by a single
  size of $a$ (spherical radius) with a number density
  of $n_{\rm d}$. The visual extinction caused by these
  grains with a pathlength of $L$ is
  $A_V = 1.086\,\pi a^2 Q_{\rm ext}(V)\,n_{\rm d} L$,
  where $Q_{\rm ext}(V)$ is the dust extinction efficiency at 
  $V$-band ($\lambda=5500$\,\AA).
  The dust number density can be derived from 
  \begin{equation}
  n_{\rm dust} \approx \frac{\langle A_V/L\rangle} 
                 {1.086\,\pi a^2 Q_{\rm ext}(V)}
                 \approx 1.1\times 10^{-12}
  \,\left(\frac{\langle A_V/L\rangle}{1.8\magkpc}\right) 
              \left(\frac{1.5}{Q_{\rm ext}[V]}\right)
 \left(\frac{0.1\mum}{a}\right)^2~~.
 \end{equation}
  The dust mass density is approximately
  $\rho_{\rm dust} = n_{\rm dust} \left(4/3\right)\,\pi a^3\rho_{\rm d}
  \approx 1.2\times 10^{-16}\g\cm^{-3}$ if we take $a\approx 0.1\mum$,
  $Q_{\rm ext}(V)=1.5$, and $\rho_{\rm d}=2.5\g\cm^{-3}$. 
  }
\\[2mm] 
The association of interstellar dust and gas had been
demonstrated by Bohlin, Savage, \& Drake (1978) who found that
the color excess and the total hydrogen column density 
(determined from the observations of HI Lyman-$\alpha$ and 
H$_2$ absorption lines with the {\it Copernicus} satellite) 
were well correlated:
$E(B-V)/\NH\approx 1.7\times 10^{-22}\magni\cm^2$
for the diffuse ISM in the solar neighbourhood. 
This correlation has recently been confirmed by
the observations with 
the {\it Far Ultraviolet Spectroscopic Explorer}
(FUSE) up to $E(B-V)\approx 1.0$ 
(Rachford et al.\ 2002),\footnote{%
  Dark clouds (e.g. the $\rho$ Oph cloud) seem to have lower 
  $E(B-V)/\NH$ values, suggesting grain growth through
  coagulation (Jura 1980; Vrba, Coyne, \& Tapia 1993;
  Kim \& Martin 1996).
  } 
suggesting that {\bf the dust and gas 
are generally well mixed in the ISM}.
From this ratio of $E(B-V)$ to $\NH$ one can estimate 
the gas-to-dust mass ratio to be $\simali$210 in 
the diffuse ISM if we take $R_V\approx 3.1$ (see Footnote-2
in Li 2004b); together with the ``rate of extinction''
$\langle A_V/L\rangle \approx 1.8\magkpc$, one can
estimate the hydrogen number density to be
$n_{\rm H} = R_V\,\langle A_V/L\rangle\,\NH/E(B-V) \approx 1.1\cm^{-3}$ 
and a gas mass density of 
$\rho_{\rm gas} \approx 2.6\times 10^{-24}\g\cm^{-3}$.
\end{itemize}

\ack I thank the organizers F. Borghese and R. Saija
for inviting me to this very exciting and fruitful conference.
I thank F. Borghese, C. Cecchi-Pestellini, A. Giusto,
M.A. Iat\`i, M.I. Mishchenko, and R. Saija for helpful discussions.


\end{document}